\documentclass{PoS}

\usepackage{amssymb}
\usepackage{graphicx}
\usepackage{epsfig}

\title{Collapse of Flow: Probing the Order of the Phase Transition}

\ShortTitle{Collapse of Flow}

\author{\speaker{Horst~St\"ocker}\\
        FIAS- Frankfurt Institute for Advanced Studies,\\
        Max-von-Laue-Str.~1, 60438 Frankfurt, Germany,\\
        Institut f\"ur Theoretische Physik, Johann Wolfgang
        Goethe - Universit\"at,\\
        Max-von-Laue-Str.~1, 60438 Frankfurt, Germany\\
	Gesellschaft f\"ur Schwerionenforschung (GSI),\\
	Planckstr. 1, 64291 Darmstadt \\
        E-mail: \email{stoecker@fias.uni-frankfurt.de}}

\abstract{We discuss the present collective flow signals for the phase
  transition to the quark-gluon plasma (QGP) and the collective flow as a
  barometer for the equation of state (EoS). We emphasize the
importance of the flow excitation function from $1$ to $50 A$~GeV:
here the hydrodynamic model has predicted the collapse of the
$v_1$-flow at $\sim 10 A$~GeV and of the $v_2$-flow at $\sim 40 A$~GeV. 
In the latter case, this has recently been observed by the NA49 collaboration.
Since hadronic rescattering models predict much larger flow than
observed at this energy, we interpret this observation as potential
evidence for a first order phase transition at high baryon density
$\rho_B$.}

\FullConference{The 4rd edition of the International Workshop --- The
  Critical Point and Onset of Deconfinement --- \\
		 July 9-13 2007\\
		 Gesellschaft f\"ur Schwerionenforschung, Darmstadt, Germany}

\begin{document}

\section{The QCD phase diagram}

\begin{figure}[t]
\centerline{\epsfig{file=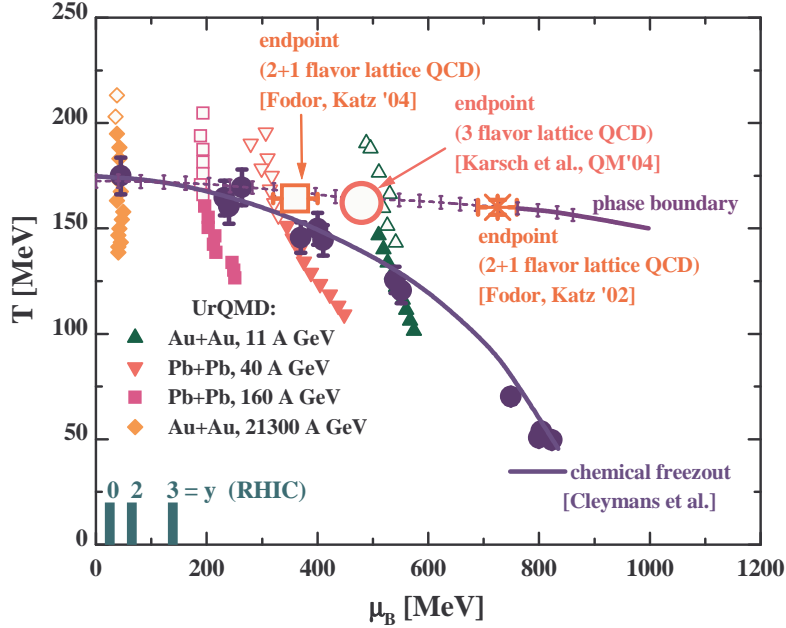,scale=0.55}}
\caption[]{The phase diagram with the critical end point at $\mu_B
  \approx 400 \mbox{ MeV}, T \approx 160 \mbox{ MeV} $, predicted by
  Lattice QCD calculations. For different bombarding energies, 
   the time evolution in the $T-\mu_B$--plane of a central cell in UrQMD calculations 
   \cite{Bravina} is depicted. (from Bratkovskaya {\it
    et al.})\protect{~\cite{Bratkov04}}.}
\label{phasedia}
\end{figure}

The phase diagram predicted by lattice QCD calculations 
\cite{Fodor04,Karsch04} (Fig.~\ref{phasedia}) shows a
cross over for vanishing or small chemical potentials $\mu_B$,
but no first-order phase transition to the  quark-gluon
plasma (QGP). This region may be accessible at full RHIC
energy. In contrast, at lower SPS and
RHIC energies ($\sqrt{s}\approx 4-12 A$~GeV) and in the fragmentation
region of RHIC, $y \approx 3-5$ \cite{Anishetty80,Date85}
a first-order phase transition is expected with a
critical baryochemical potential of \cite{Fodor04,Karsch04}
$\mu_B^c \approx 400 \pm 50 \mbox{ MeV}$ and a
critical temperature of $T_c \approx 150-160$ MeV. This first-order 
phase transition is expected to occur at finite
strangeness~\cite{Greiner:1987tg}. 

A comparison of the QCD predictions of the thermodynamic parameters
$T$ and $\mu_B$ with the results from the UrQMD transport model 
\cite{Bass:1998ca,Bleicher:1999xi} in the central overlap
regime of Au+Au collisions~\cite{Bratkov04} are shown in 
Figure~\ref{phasedia}. The 'experimental' chemical freeze-out parameters --
determined from fits to the experimental yields -- are shown by full dots
with errorbars and taken from Ref.~\cite{Cleymans}. 
The temperature $T$ and chemical potentials
$\mu_B$, denoted by triangular and quadratic symbols (time-ordered in
vertical sequence), are taken from UrQMD transport calculations in central Au+Au
(Pb+Pb) collisions at RHIC~\cite{Bravina} as a function of the reaction
time (separated by 1 fm/c steps from top to bottom). 
Full symbols denote configurations in approximate pressure equilibrium in
longitudinal and transverse direction, while open symbols
denote nonequilibrium configurations and correspond to $T$ parameters
extracted from the transverse momentum distributions.

The transport calculations during the nonequilibrium phase (open symbols) 
show much higher temperatures (or energy densities) than
the 'experimental' chemical freeze-out configurations at all bombarding
energies ($\geq 11 A$~GeV). These numbers exceed the critical point 
of (2+1) flavor lattice QCD calculations
by the Bielefeld-Swansea-collaboration~\cite{Karsch04} (large open
circle) and by the Wuppertal-Budapest-collaboration~\cite{Fodor04} (open square; the
star denotes earlier results from~\cite{Fodor04}). The energy density at
$\mu_c, T_c$ is of the order of $\approx$ 1 GeV/fm$^3$. 
At RHIC energies, when the temperature drops during the expansion phase of the 'hot
fireball' a cross over is expected at midrapidity. 
Using the statistical model analysis by the BRAHMS collaboration based on
measured antibaryon to baryon ratios~\cite{BRAHMS_PRL03}  for different
rapidity intervals at RHIC energies, the 
baryochemical potential $\mu_B$ has been obtained. At midrapidity, one observes
$\mu_B\simeq 0$, whereas at forward rapidities $\mu_B$ increases up to
$\mu_B\simeq 130$~MeV at $y=3$. Thus, only a forward rapidity
measurement ($y \approx 4-5)$ at RHIC will allow to probe large $\mu_B$. 
A unique opportunity to reach higher chemical potentials and the first-order phase
transition region at midrapidity is offered by  
the STAR and PHENIX detectors at RHIC in the high-$\mu$-RHIC-running at
$\sqrt{s}=4-12 A$~GeV. For first results see Ref.\ \cite{RoehrichBrahms}. The International FAIR
Facility at GSI will offer a research program fully devoted to this topic in
the next decade.

\subsection{Flow Effects from Hydrodynamics}

Early in the 70th, hydrodynamic flow and shock formation have been proposed
~\cite{Hofmann74,Hofmann76} as the key mechanism for the creation
of hot and dense matter in relativistic heavy-ion
collisions~\cite{Lacey}. Though, the full three-dimensional hydrodynamical flow
problem is much more complicated than the one-dimensional Landau
model~\cite{Landau}. The 3-dimensional compression and expansion
dynamics yields complex triple differential cross sections which
provide quite accurate spectroscopic handles on the EoS. 
Differential barometers for the properties of compressed, dense matter from SIS to
RHIC are the bounce-off, $v_1(p_T)$ (i.e., the strength of the directed flow in the
reaction plane), the squeeze-out, $v_2(p_T)$ (the strength of the
second moment of the azimuthal particle emission
distribution)~\cite{Hofmann74,Hofmann76,Stocker79,Stocker80,Stocker81,Stocker82,Stocker86},
and the antiflow~\cite{Stocker79,Stocker80,Stocker81,Stocker82,Stocker86}
(third flow component~\cite{Csernai99,Csernai04}). It has been
shown~\cite{Hofmann76,Stocker79,Stocker80,Stocker81,Stocker82,Stocker86}
that the disappearance or so-called collapse of flow is a direct result of a
first-order phase transition.

To determine these different barometers, several hydrodynamic models~\cite{Rischke:1995pe} 
have been used in the past, starting with the one-fluid ideal hydrodynamic approach. 
It is known that this model predicts far too large flow
effects so that viscous fluid models have been developed~\cite{Schmidt93,Muronga01,Muronga03}
to obtain a better description of the dynamics. In
parallel, so-called three-fluid models, which distinguish between
projectile, target and the fireball fluid, have been
considered~\cite{Brachmann97}. Here viscosity effects do not appear 
inside the individual fluids, but only between different fluids.
One aim is to obtain a reliable,
three-dimensional, relativistic three-fluid model including
viscosity~\cite{Muronga01,Muronga03}.

Though flow can be described very elegantly in hydrodynamics, one should 
consider microscopic multicomponent (pre-)hadron transport theory,
e.g. models like qMD~\cite{Hofmann99}, IQMD~\cite{Hartnack89}, 
UrQMD~\cite{Bass:1998ca,Bleicher:1999xi}, or HSD~\cite{Cassing99}, to control models for viscous
hydrodynamics and to gain background models to subtract interesting
non-hadronic effects from data. If hydrodynamics with and without
quark matter EoS and hadronic transport models without quark matter --
but with strings -- are compared to data, can we learn whether quark
matter has been formed? What degree of equilibration has been
reached? What does the EoS look like? How are the
particle properties, self-energies, cross sections changed?

\subsection{Evidence for a first--order phase transition from AGS and SPS}

\begin{figure}[t]
\begin{minipage}[r]{5 cm}
\hspace{1cm}
\epsfig{file=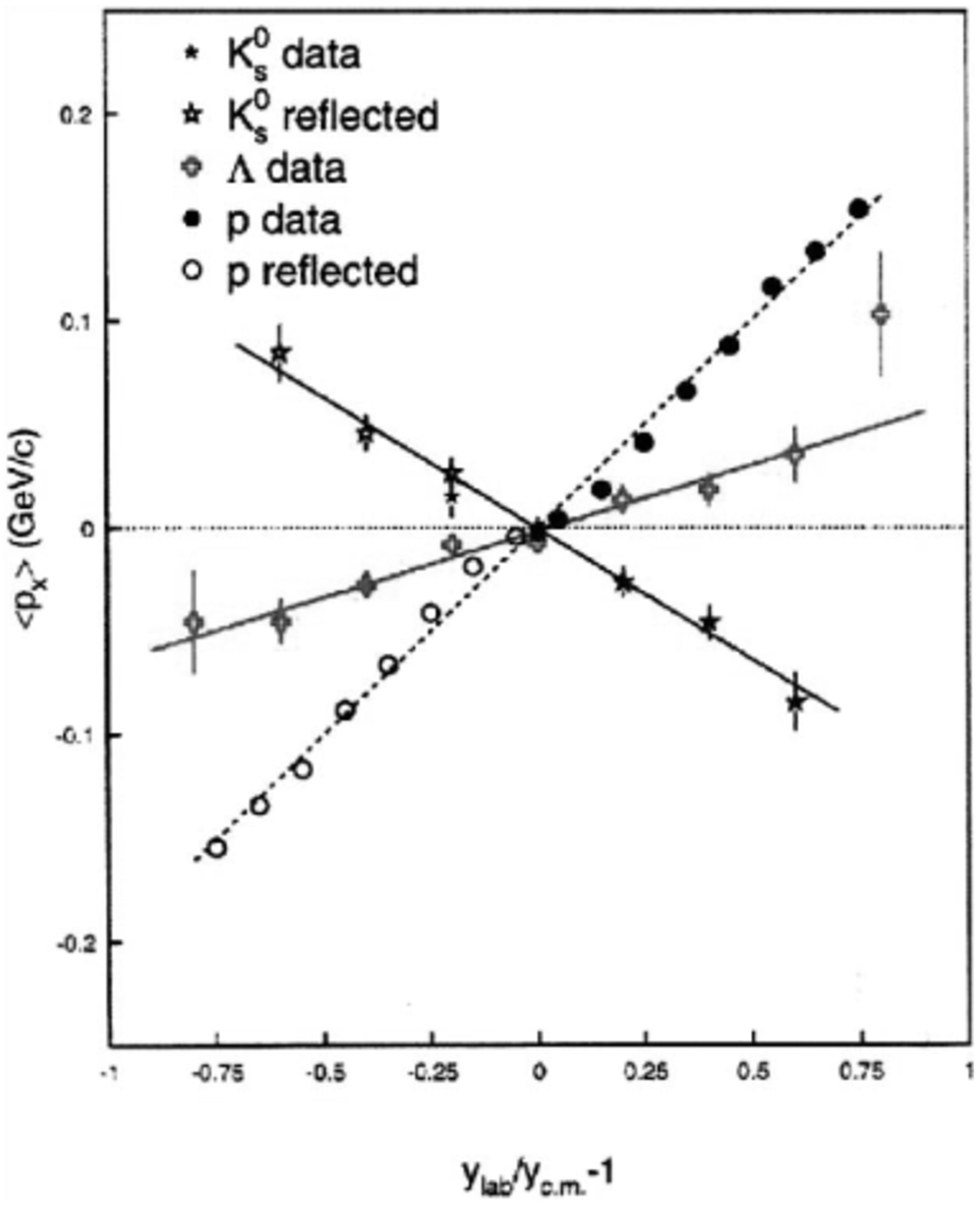,scale=0.4}
\end{minipage}
\begin{minipage}[l]{8 cm}
\vspace{-0.8cm}
\hspace{1cm}
\epsfig{file=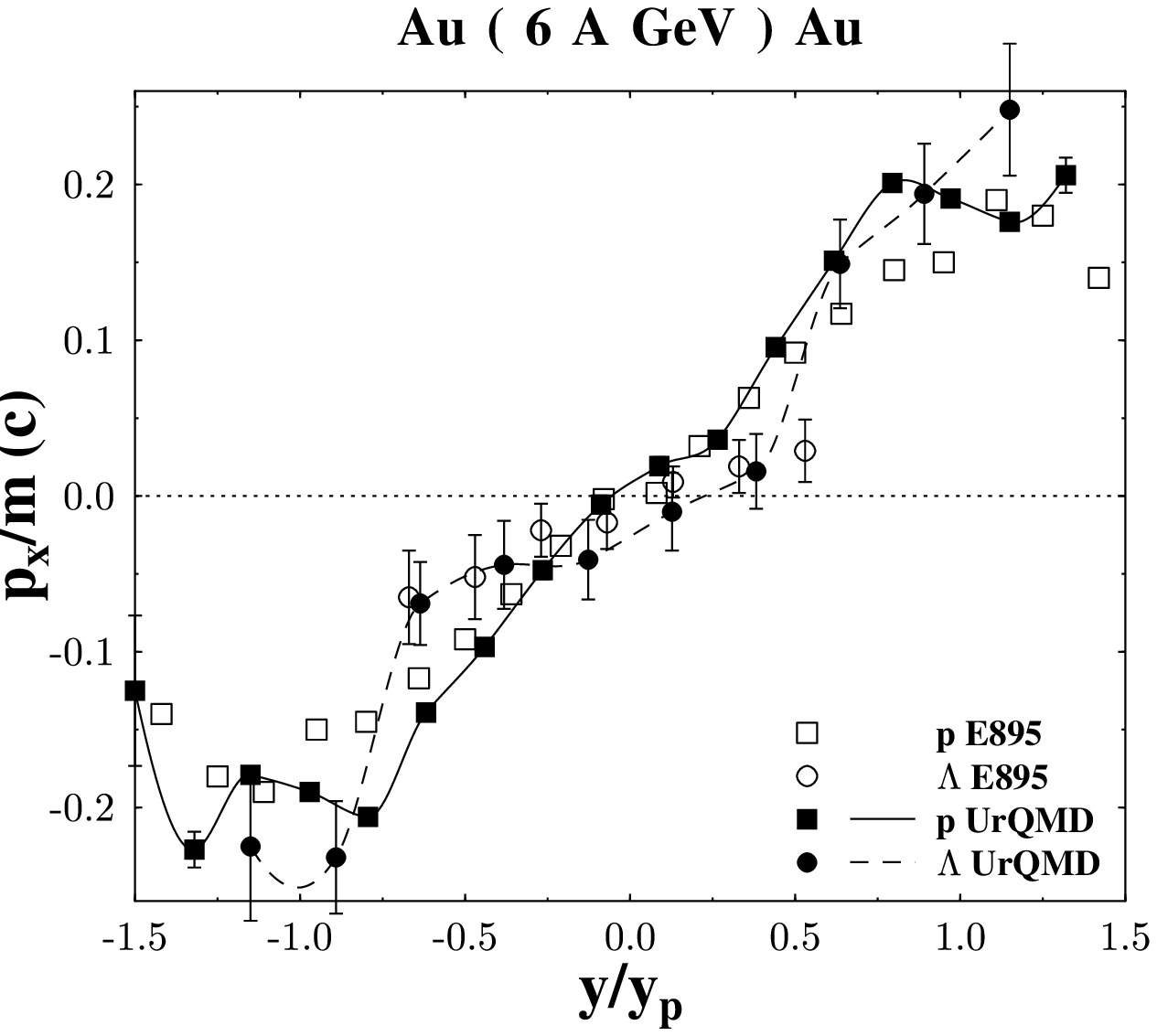,scale=0.6}
\end{minipage}
\caption{
Sideward flow $p_x$ of (left) K, $\Lambda$ and p's
at $6 A$~GeV as measured by E895 in semi-central collisions at the AGS
and (right) for $p$ and $\Lambda$ compared to UrQMD1.1 calculations for $b < 7$ fm
\protect{\cite{Soff99} }. }
\label{flow_ags_soff}
\end{figure}

\begin{figure}[t]
\vspace*{0.5cm}
\centerline{\epsfig{file=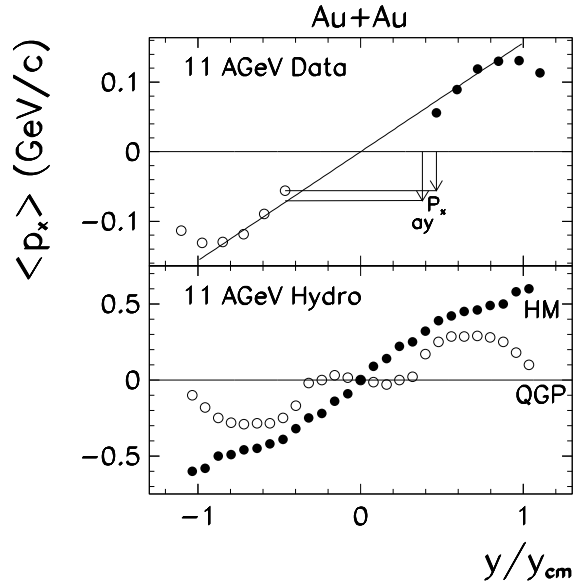,scale=0.45}}
\caption{Prediction of the directed flow from 
ideal hydrodynamics with  a QGP phase (open symbols)
and from the Quark Gluon String Model without QGP phase (full symbols)
\protect{\cite{Csernai99} }. \label{flow_csernai}}
\end{figure}

\begin{figure}[t]
\epsfig{file=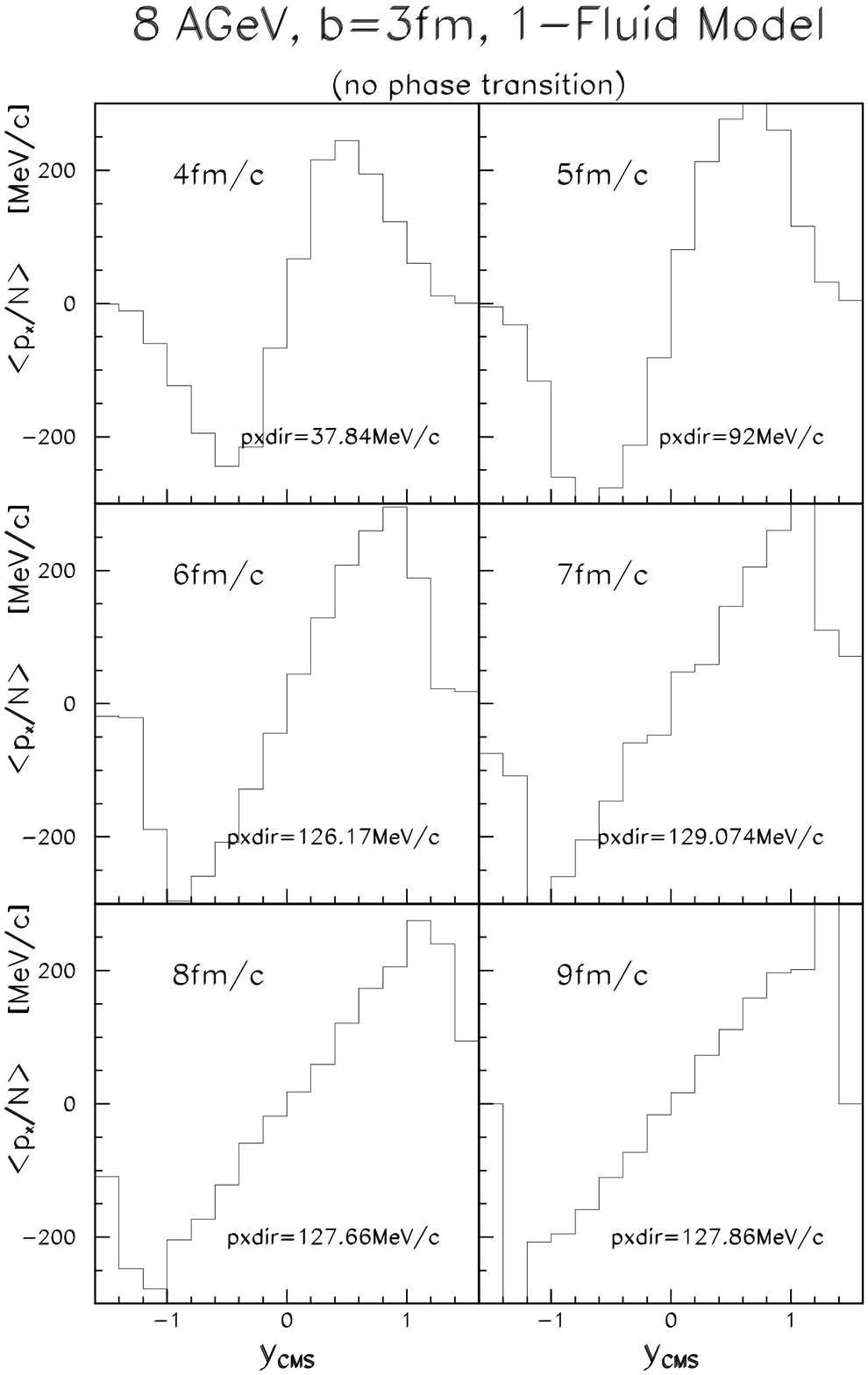,width=6.2cm}\hspace*{1.5mm}
\epsfig{file=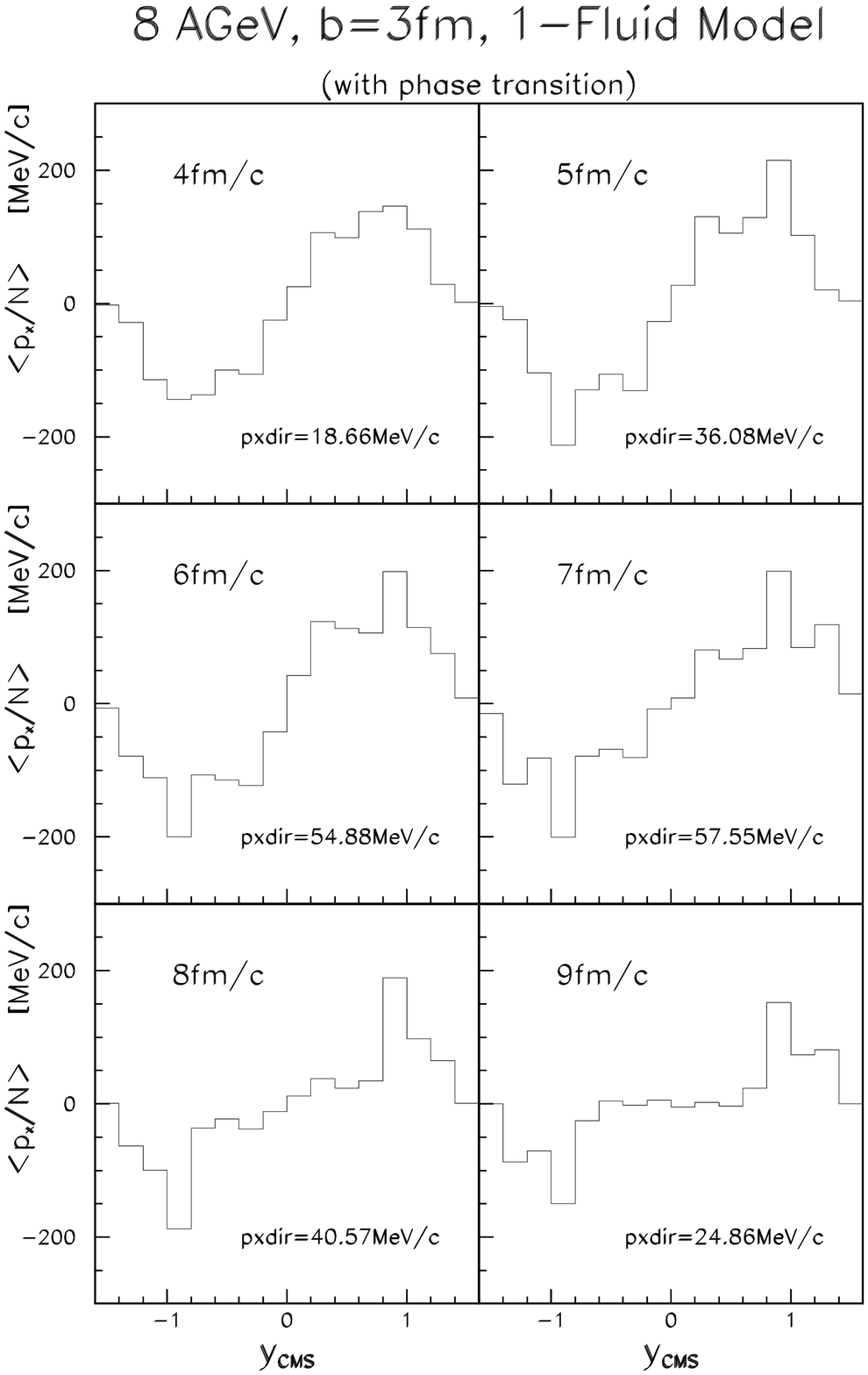,width=6.2cm}
\caption[]{The time evolution of directed flow $p_x/N$ as a function of
  rapidity for Au+Au collisions at $8 A$~GeV in the one-fluid
  model for (left) a hadronic EoS without phase transition 
  and (right) an EoS including a first-order phase transition to the QGP [from
  Brachmann]\protect{\cite{Brach00}}.
\label{flow_brach1}} 
\end{figure}

\begin{figure}[t]
\centerline{\epsfig{file=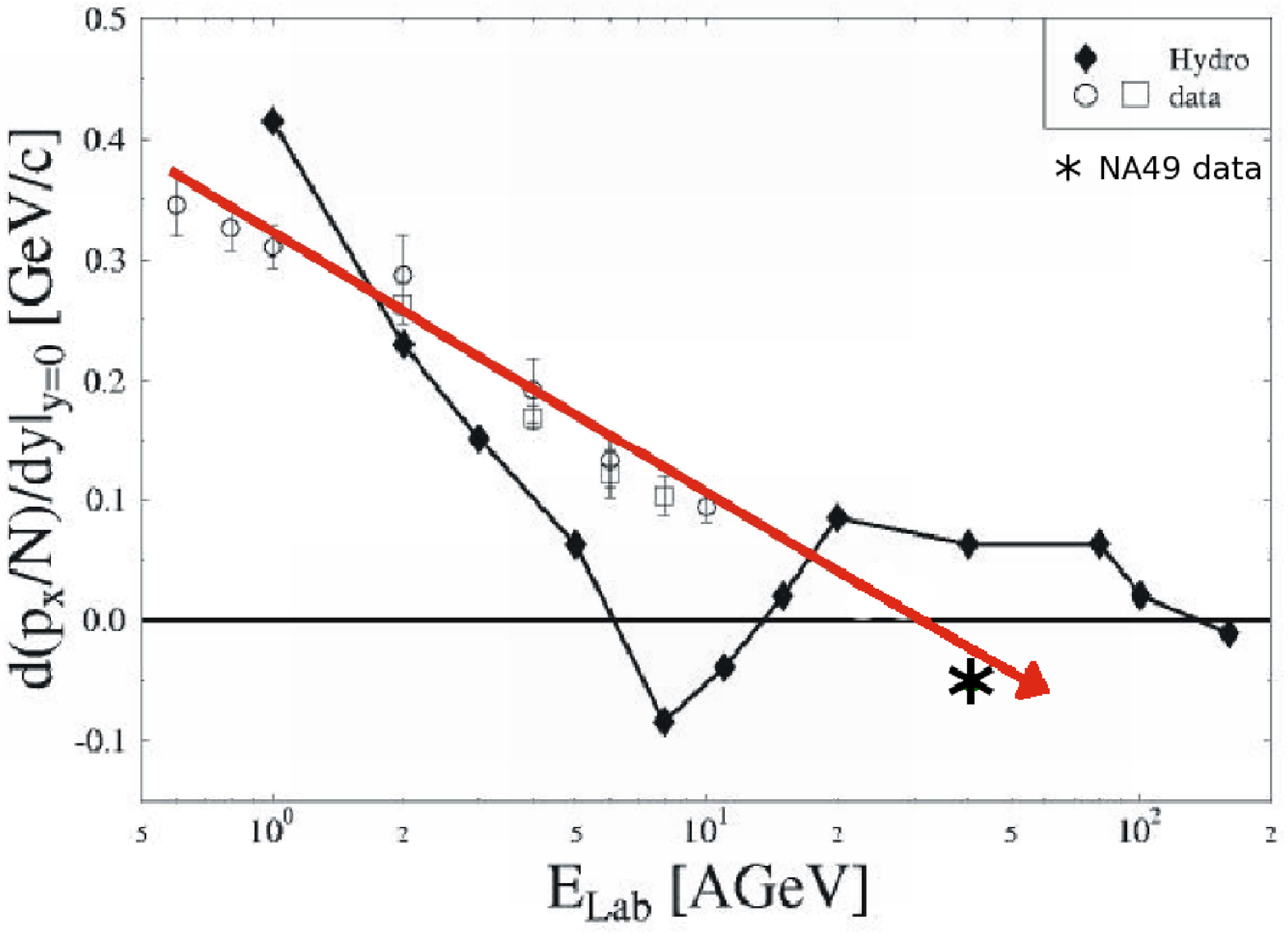,scale=0.40}}
\caption[]{The proton $dp_x/dy$-slope data measured by SIS and AGS compared
  to a one-fluid hydrodynamical calculation. A linear extrapolation of
  the AGS data indicates a collapse of flow at $E_{Lab} \approx 30 A$~GeV 
  (see also Ref.\protect{~\cite{Brach99}}). The point at $40 A$~GeV is
    calculated using the NA49 central data (cf. Alt {\it et
    al.})\protect{~\cite{NA49_v2pr40}}.}
\label{flow_extra}
\end{figure}

The formation and distribution of many hadronic particles at AGS and
SPS is quite well described by microscopic (pre-)hadronic transport models
\cite{Weber02}. Additionally, flow data 
are described reasonably well up to AGS
energies~\cite{Csernai99,Andronic03,Andronic01,Soff99,Sahu1,Sahu2}, 
if a nuclear potential has been included for the low energy regime. 

However, since ideal hydrodynamical calculations predict far too
much flow at these energies~\cite{Schmidt93}, viscosity
effects have to be taken into account. While the
directed flow $p_x/m$ measurement of the E895 collaboration shows that
the $p$ and $\Lambda$ data are reproduced reasonably 
well~\cite{Soff99,Sto04}, ideal hydrodynamical calculations yield factors of two
higher values for the sideward flow at SIS~\cite{Schmidt93} and AGS.

However, the appearance of a so-called ''third flow component'' \cite{Csernai99} or ''antiflow''
\cite{Brach00} in central collisions (cf. Fig. \ref{flow_csernai}) is 
predicted in ideal hydrodynamics, though only if the matter undergoes a
first order phase transition to the QGP. It implies that around midrapidity
the directed flow, $p_x (y)$, of protons develops a negative slope. Such an
exotic ''antiflow'' (negative slope) wiggle in the proton flow
$v_1(y)$ does not appear for a hadronic EoS without QGP phase transition at intermediate energies.
For high energies see disussion in References \cite{Snellings:1999bt,Bleicher:2000sx}.
Just as the microscopic transport theory (Fig. \ref{flow_ags_soff} r.h.s.) and as the data
(Fig.  \ref{flow_ags_soff} l.h.s.), the ideal hydrodynamic time evolution of the directed flow,
$p_x/N$, for the purely hadronic EoS (Fig.  \ref{flow_brach1} l.h.s.)
does show a clean linear increase of $p_x(y)$. However, it can be seen that for an EoS
including a first order phase transition to the QGP (Fig.  \ref{flow_brach1} r.h.s.)
that the proton flow $v_1 \sim p_x/p_T$ collapses around midrapidity. 
This is explained by an antiflow component of protons that develops when
the expansion from the plasma sets in \cite{Brach99}. 

Even negative values of $d(p_x/N)/dy$ calculated from
ideal hydrodynamics (Fig. \ref{flow_extra}) show up between $8$ and $20
A$~GeV. An increase up to positive values is predicted with
increasing energy. But, the hydro
calculations suggest this ''softest point collapse'' is at $E_{Lab}
\approx 8  A$~GeV. This predicted minimum of the proton flow 
has not been verified by the AGS data! However, a collapse of
the directed proton flow at $E_{Lab} \approx 30 A$~GeV (Fig.
\ref{flow_extra}) is verified by a linear extrapolation of the AGS data. 

\begin{figure}[t]
\begin{center}
\begin{minipage}[r]{5 cm}
\hspace{1cm}
\epsfig{file=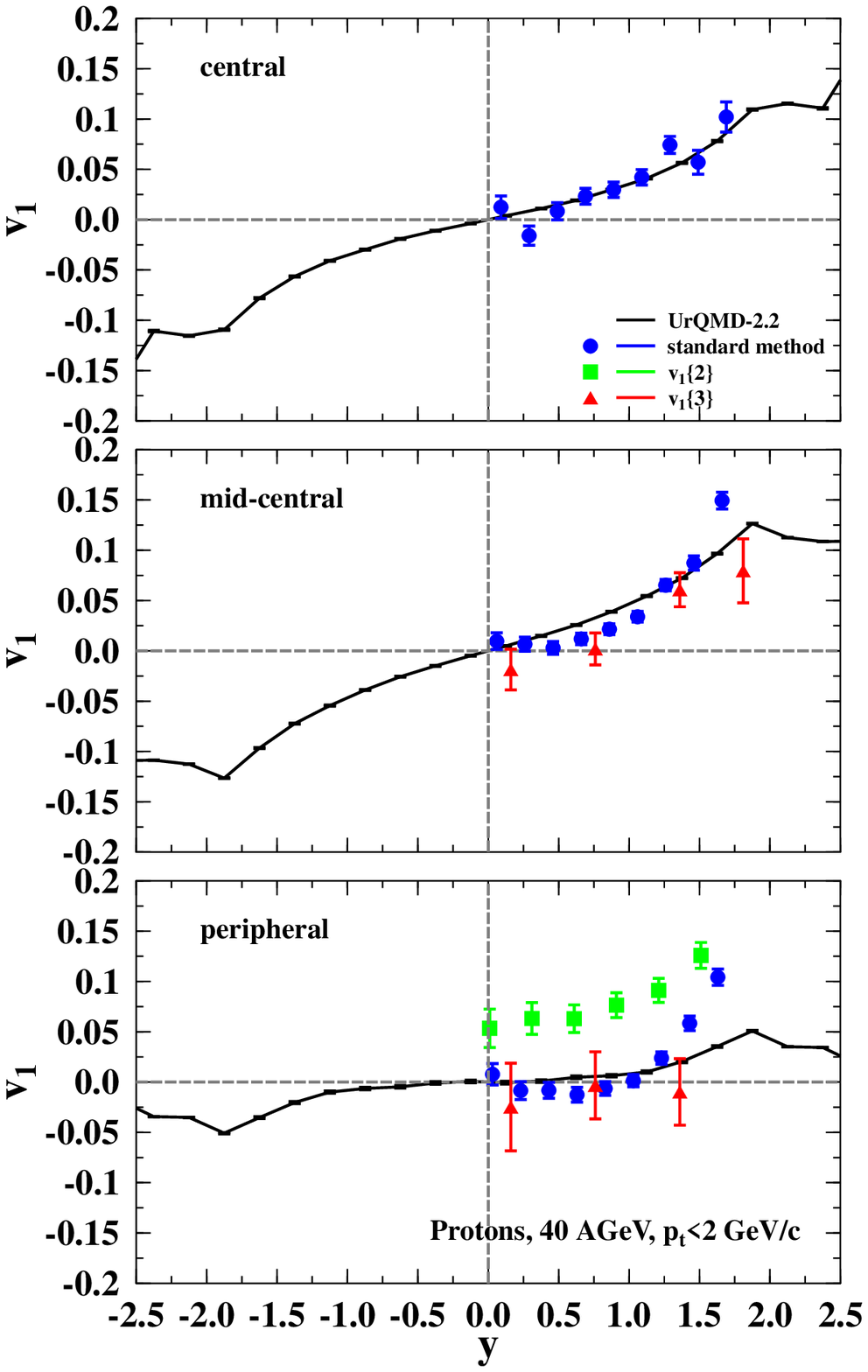,scale=0.4}
\end{minipage}
\begin{minipage}[r]{5 cm}
\hspace{1cm}
\epsfig{file=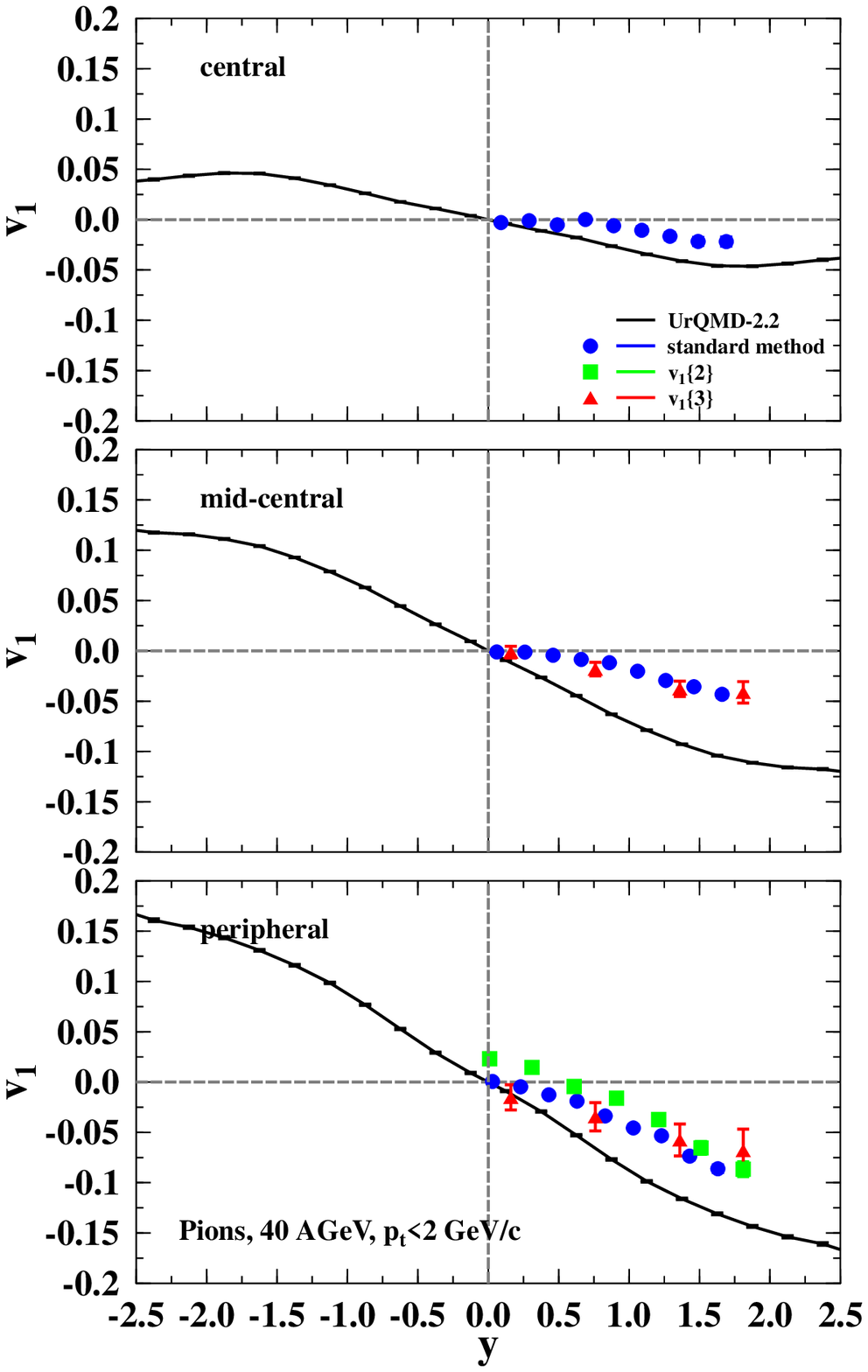,scale=0.4}
\end{minipage}
\caption[]{(Color online)Directed flow of protons (left) and pions (right) in Pb+Pb collisions at 
$E_{\rm lab}=40A~$GeV with $p_{t} < 2$~GeV/c. UrQMD 
calculations are depicted with black lines. The symbols are NA49 data from different analysis methods. The standard 
method (circles), cumulant method of order 2 (squares) and cumulant method of order 3 (triangles) are depicted. The 
12.5\% most central collisions are labeled as central, the centrality 12.5\% -33.5\% as mid-central and 33.5\% -100\% 
as peripheral. For the model calculations the corresponding impact parameters of $b \le 3.4$~fm for central, 
$b=5-9$~fm for mid-central and $b= 9-15$~fm for peripheral collisions have been used 
(from Petersen {\it et. al.} \protect{\cite{Petersen}}).} 
\label{Hannah1}
\end{center}
\end{figure}

\begin{figure}[t]
\begin{center}
\begin{minipage}[r]{5 cm}
\hspace{1cm}
\epsfig{file=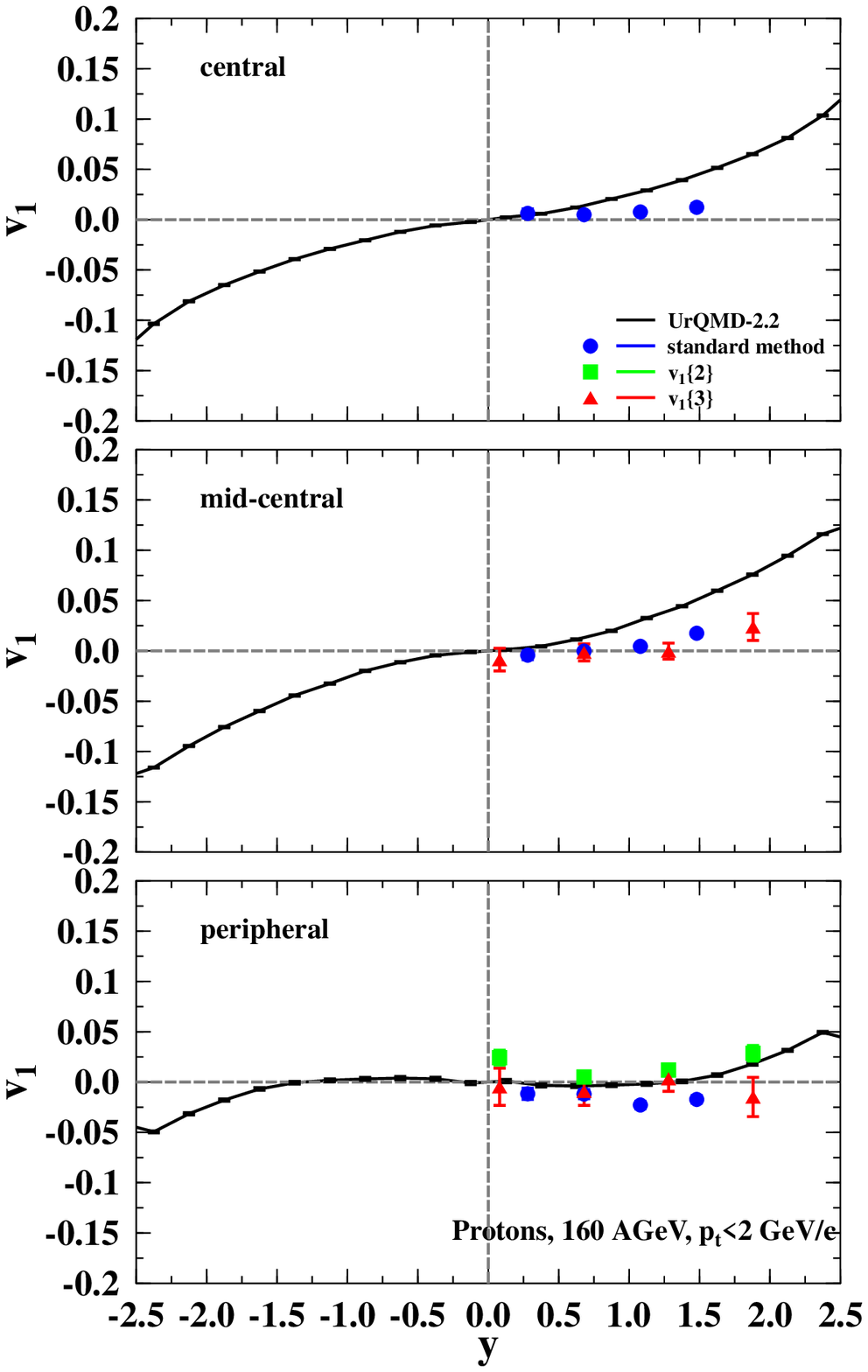,scale=0.4}
\end{minipage}
\begin{minipage}[r]{5 cm}
\hspace{1cm}
\epsfig{file=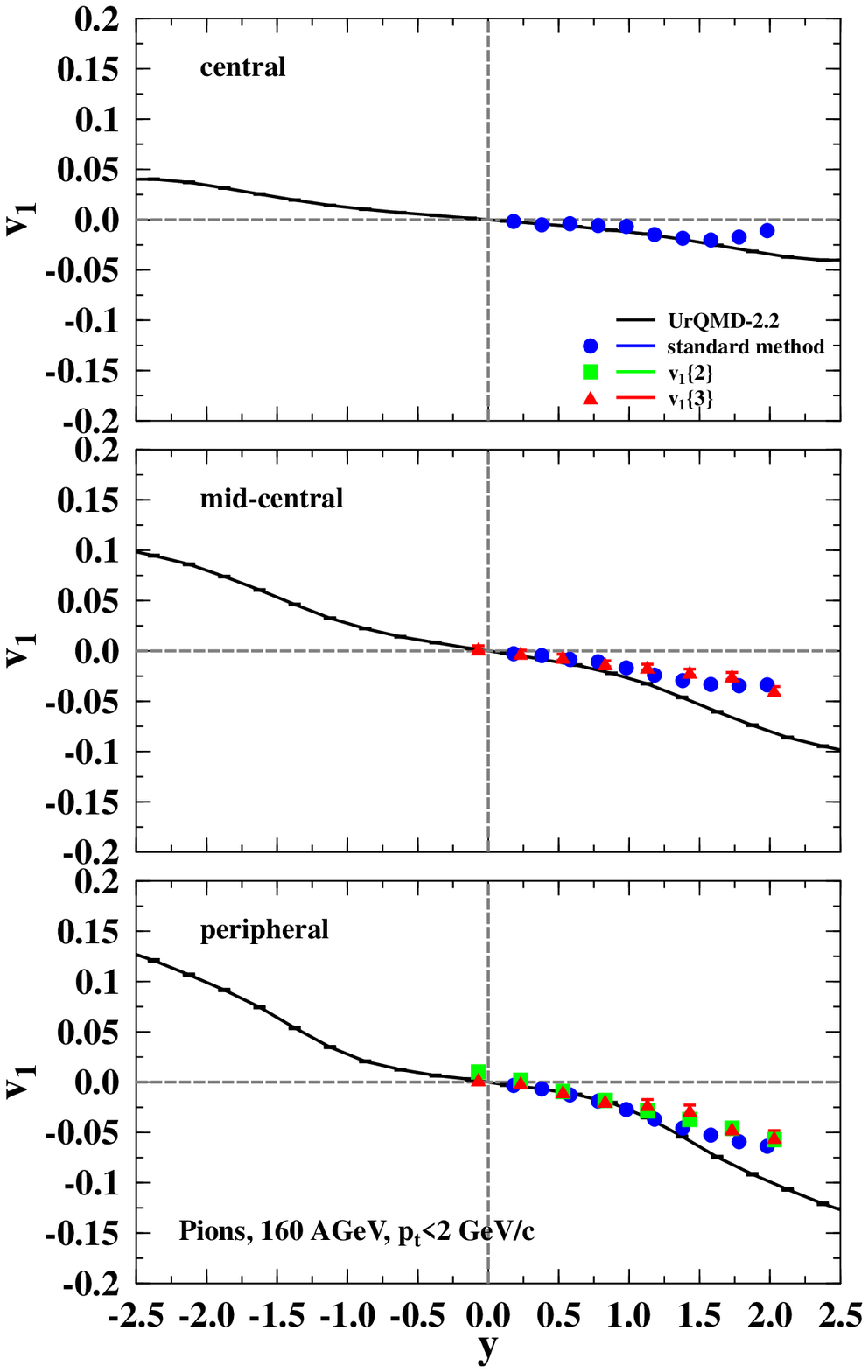,scale=0.4}
\end{minipage}
\caption[]{(Color online)Directed flow of protons (left) and pions (right) in Pb+Pb collisions at 
$E_{\rm lab}=160A~$GeV with $p_{t} < 2$~GeV/c. UrQMD 
calculations are depicted with black lines. The symbols are NA49 data from different analysis methods. The standard 
method (circles), cumulant method of order 2 (squares) and cumulant method of order 3 (triangles) are depicted. The 
12.5\% most central collisions are labeled as central, the centrality 12.5\% -33.5\% as mid-central and 33.5\% -100\% 
as peripheral. For the model calculations the corresponding impact parameters of $b \le 3.4$~fm for central, 
$b=5-9$~fm for mid-central and $b= 9-15$~fm for peripheral collisions have been used
(from Petersen {\it et. al.} \protect{\cite{Petersen}}).} 
\label{Hannah2}
\end{center}
\end{figure}

This prediction has recently been supported by the low energy
$40 A$~GeV SPS data of the NA49 collaboration
\cite{NA49_v2pr40} (cf. Figs. \ref{Hannah1} and \ref{Hannah2}). In contrast to the
AGS data as well as to the UrQMD calculations involving no phase transition 
(Figs. \ref{Hannah1} and \ref{Hannah2}), the first proton ''antiflow'' around mid-rapidity 
is clearly visible in these data. 

Thus, a first order phase transition to the baryon rich QGP is most likely observed
at bombarding energies of $30-40 A$~GeV; e.g. the first order phase
transition line in the $T$-$\mu_B$-diagram has been crossed (cf. Fig. \ref{phasedia}). 
In this energy region, the new FAIR- facility at GSI will operate. 
It can be expected that the baryon flow collapses and other first
order QGP phase transition signals can be studied soon at the lowest
SPS energies as well as at fragmentation region $y > 4-5$ for the RHIC and LHC collider energies. 
At high $\mu_B$, these experiments will enable a detailed study of the first order phase
transition as well as of the properties of the baryon rich
QGP.


\section{More evidence for a first--order phase transition at highest net baryon densities}

Microscopic transport models, at SIS energies, reproduce the data on
the excitation function of the proton elliptic flow $v_2$ quite well.
The data seem to be described well by a soft, momentum--dependent EoS
\cite{Andronic00,Andronic99}. 

Below $\sim 5 A$~GeV, the observed proton flow $v_2$ is smaller than zero, 
which corresponds to the squeeze-out predicted by hydrodynamics long
ago~\cite{Hofmann74,Hofmann76,Stocker79,Stocker80,Stocker81,Stocker82,Stocker86}.

\begin{figure}[h]
\centerline{\epsfig{file=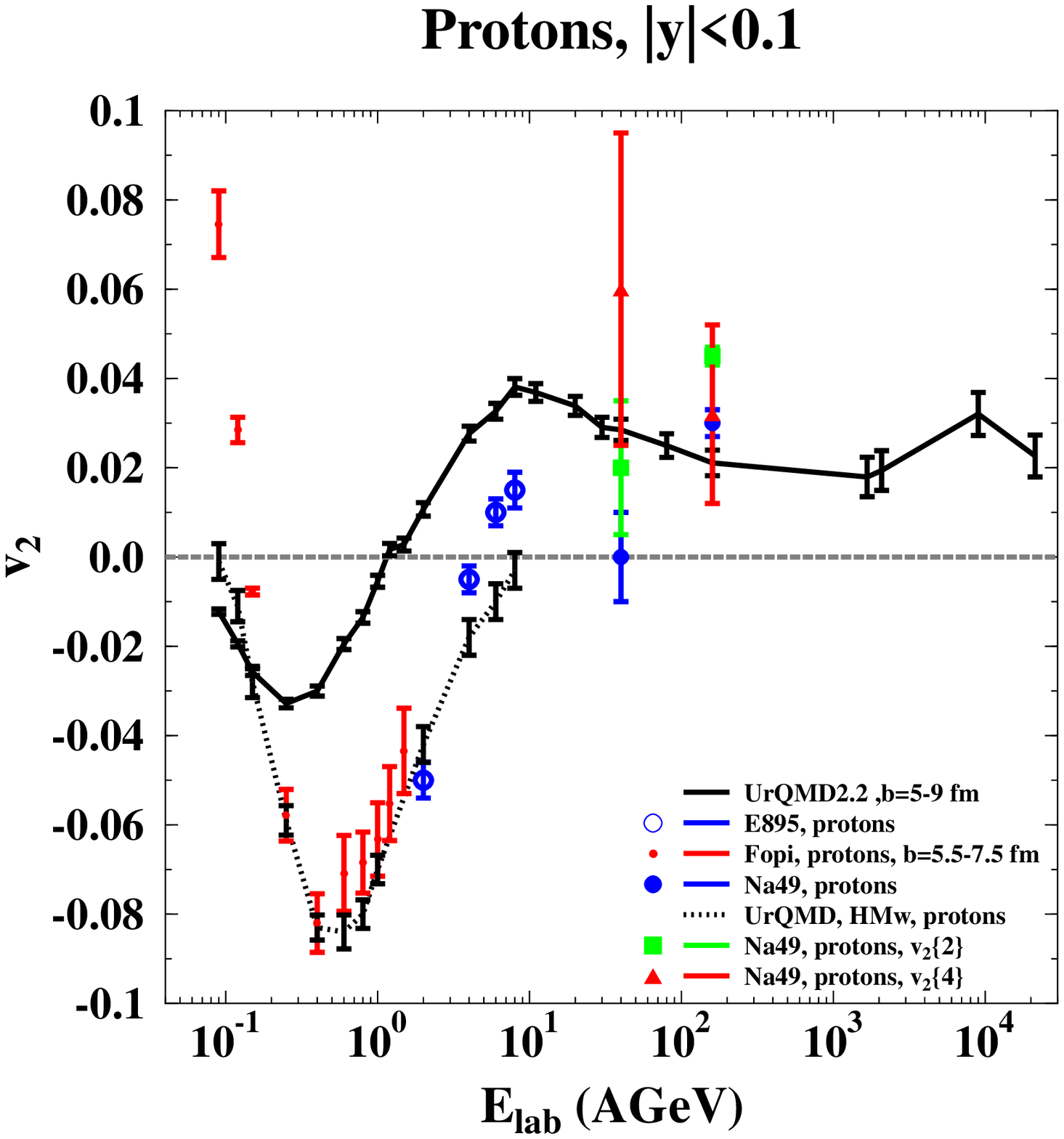,scale=0.45}}
\caption{(Color online) The calculated energy excitation function of elliptic flow of protons in Au+Au/Pb+Pb 
collisions in mid-central collisions (b=5-9 fm) with $|y|<0.1$(full line). This curve is compared to data from 
different experiments for mid-central collisions. For E895 \cite{Pinkenburg:1999nv}\cite{Chung:2001qr}, FOPI 
\cite{Andronic:2004cp} and NA49 \cite{NA49_v2pr40} there is the elliptic flow of protons. The dotted line in the 
low energy regime depicts UrQMD calculations with included nuclear potential
(from Petersen {\it et. al.} \protect{\cite{Petersen}}).}
\label{v2_excitation}
\end{figure}

From the AGS data, a transition from squeeze-out to in-plane flow in the midrapidity region can be seen
(Fig. \ref{v2_excitation}). In accord to the transport caluclations (UrQMD
calculations in Fig. \ref{v2_excitation} \cite{Soff99};
for HSD results see  \cite{Sahu1,Sahu2}), the proton  $v_2$ at $4-5 A$~
GeV changes its sign. Hadronic transport simulations predict 
a smooth increase of the flow $v_2$ at higher energies ($10-160 A$~GeV). 
The $160 A$~GeV data of the NA49 collaboration indicate that this
smooth increase proceeds as predicted between AGS and SPS. 
For midcentral and peripheral protons
at $40 A$~GeV (cf. Ref.~\cite{Soff99,Sto04}),
UrQMD calculations without phase transition give a considerable 3\% $v_2$ flow.

\begin{figure}[t]
\begin{center}
\begin{minipage}[r]{5 cm}
\hspace{1cm}
\epsfig{file=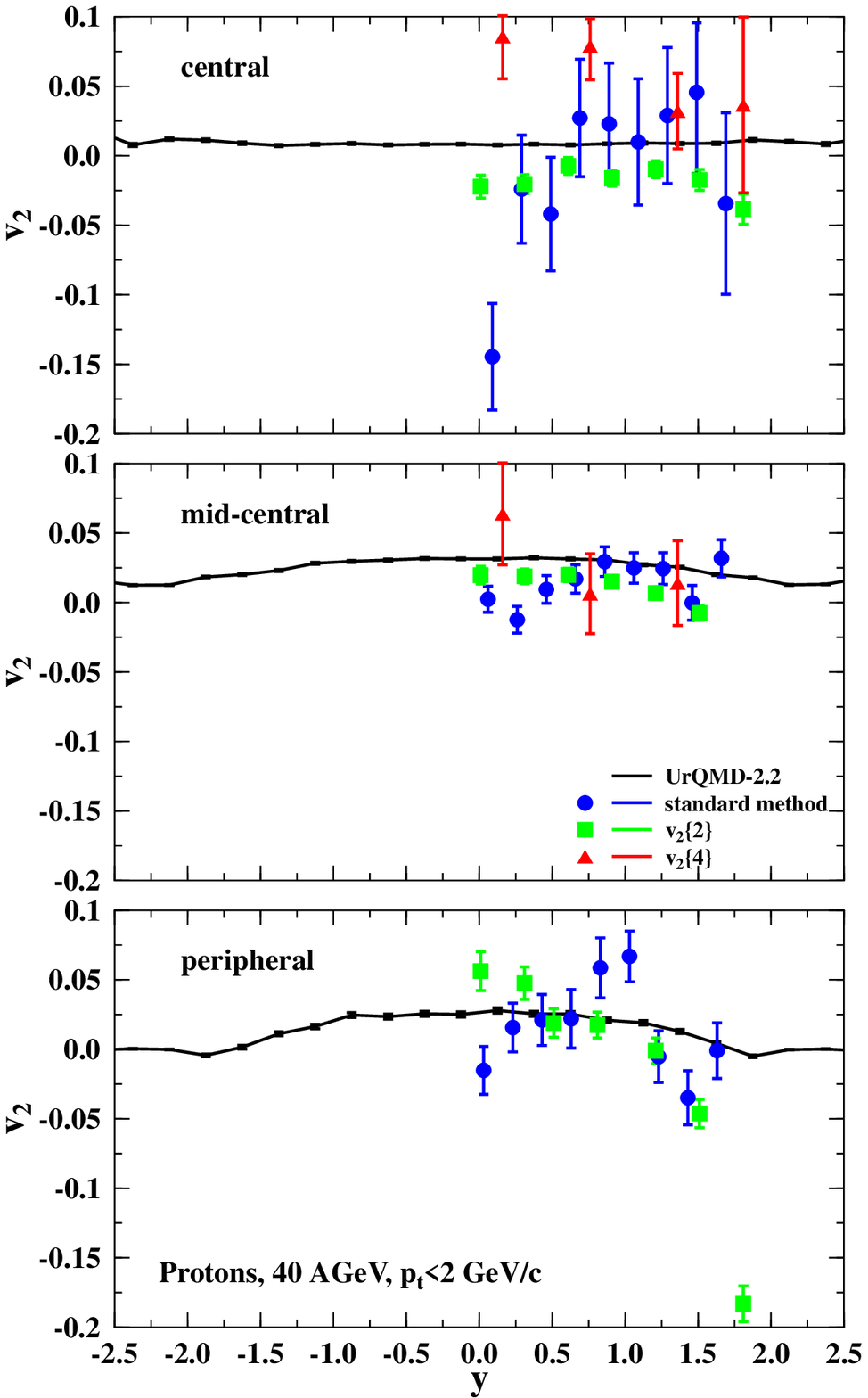,scale=0.4}
\end{minipage}
\begin{minipage}[r]{5 cm}
\hspace{1cm}
\epsfig{file=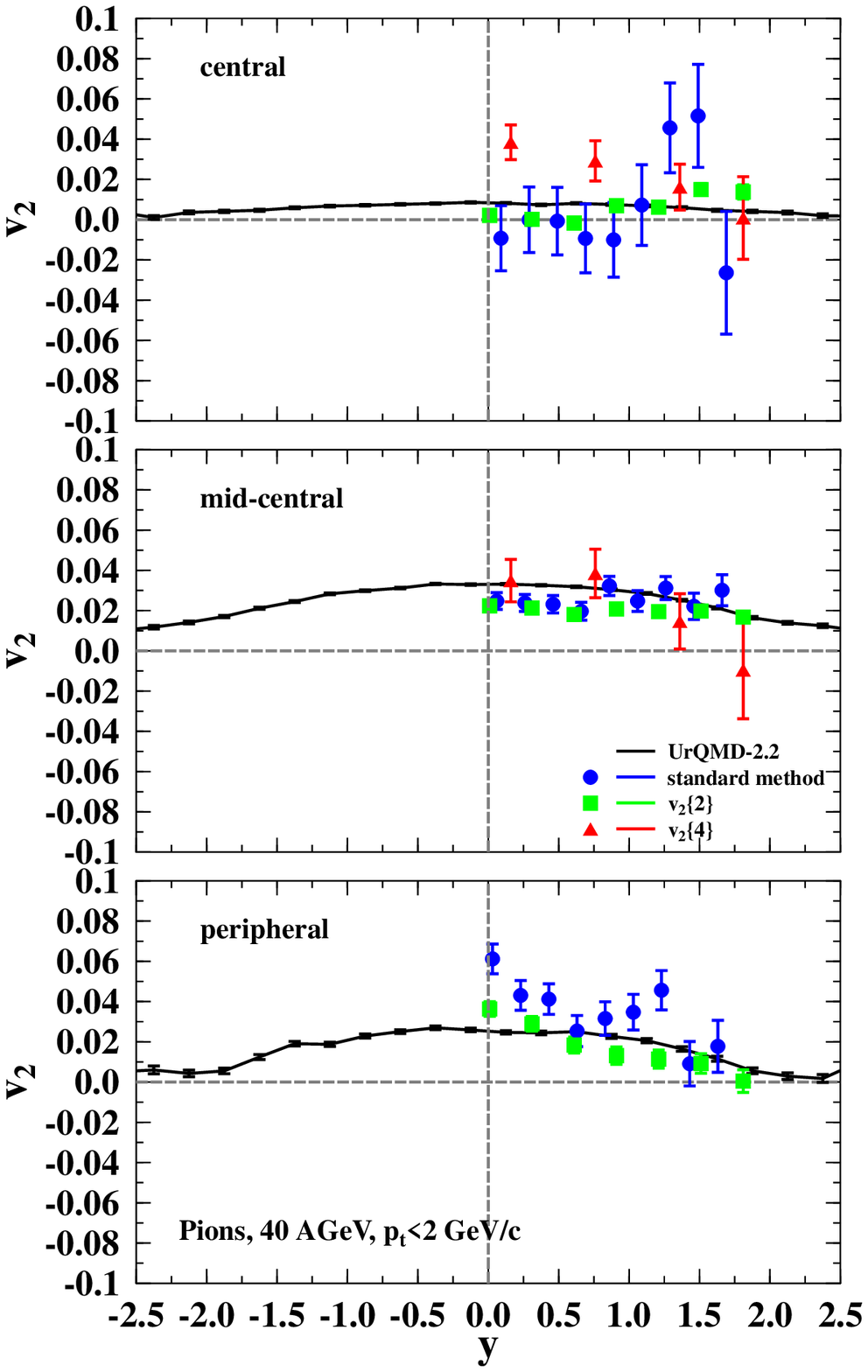,scale=0.4}
\end{minipage}
\caption[]{(Color online)Elliptic flow of protons (left) and pions (right) in Pb+Pb collisions at 
$E_{\rm lab}=40A~$GeV with $p_{t} < 2$~GeV/c. UrQMD 
calculations are depicted with black lines. The symbols are NA49 data from different analysis methods. The standard 
method (circles), cumulant method of order 2 (squares) and cumulant method of order 3 (triangles) are depicted. The 
12.5\% most central collisions are labeled as central, the centrality 12.5\% -33.5\% as mid-central and 33.5\% -100\% 
as peripheral. For the model calculations the corresponding impact parameters of $b \le 3.4$~fm for central, 
$b=5-9$~fm for mid-central and $b= 9-15$~fm for peripheral collisions have been used
(from Petersen {\it et. al.} \protect{\cite{Petersen}}).} 
\label{Hannah3}
\end{center}
\end{figure}

\begin{figure}[t]
\begin{center}
\begin{minipage}[r]{5 cm}
\hspace{1cm}
\epsfig{file=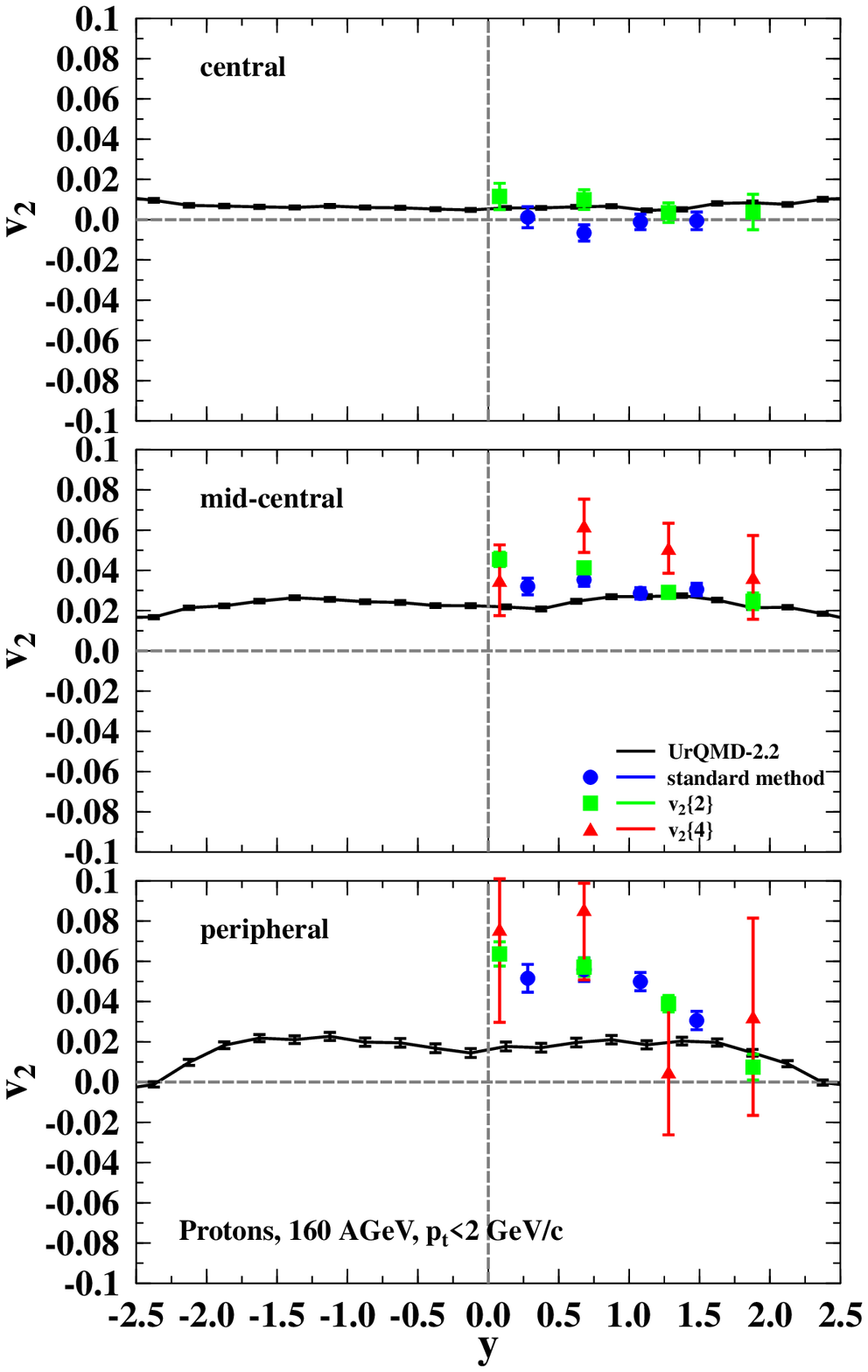,scale=0.4}
\end{minipage}
\begin{minipage}[r]{5 cm}
\hspace{1cm}
\epsfig{file=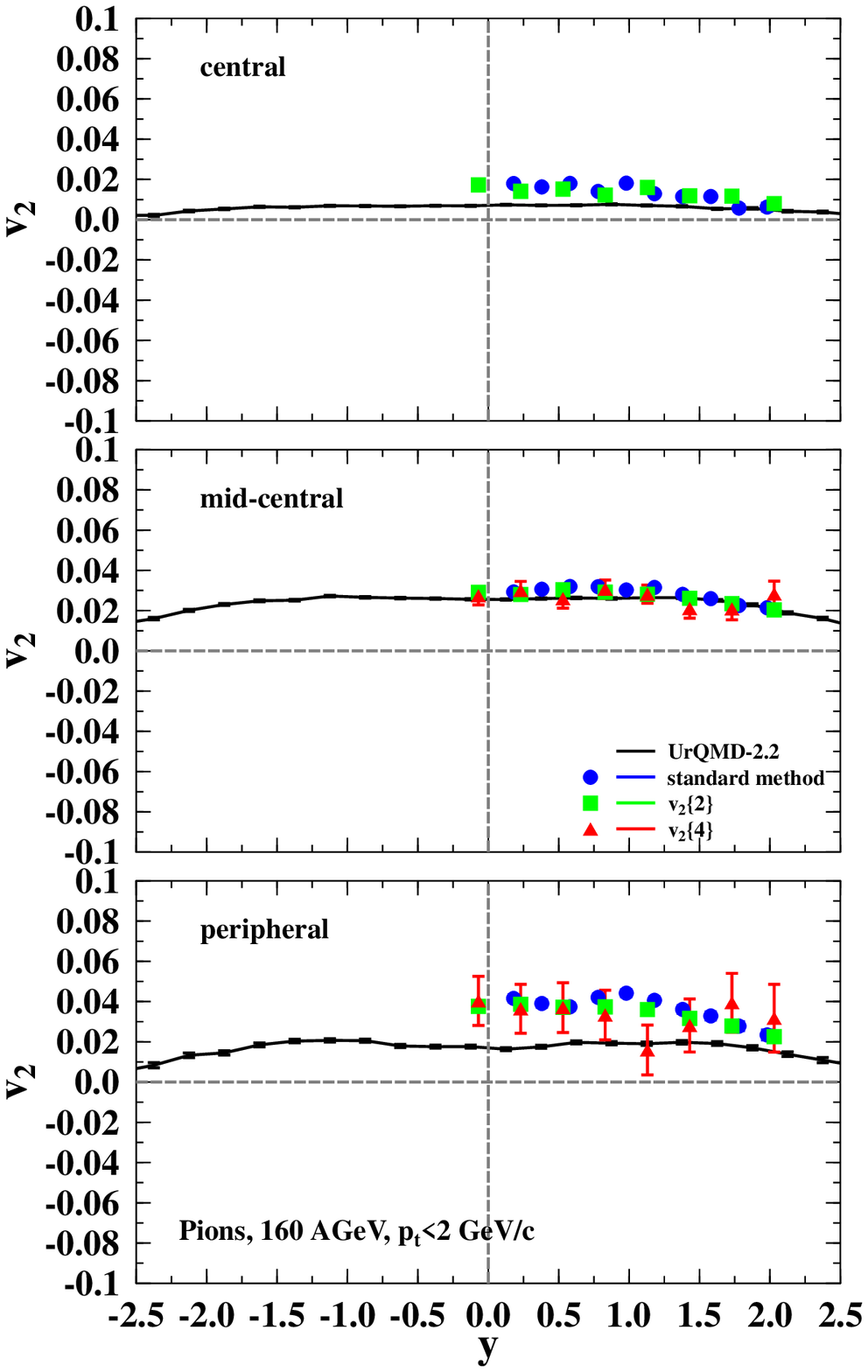,scale=0.4}
\end{minipage}
\caption[]{(Color online)Elliptic flow of protons (left) and pions (right) in Pb+Pb collisions at 
$E_{\rm lab}=160A~$GeV with $p_{t} < 2$~GeV/c. UrQMD 
calculations are depicted with black lines. The symbols are NA49 data from different analysis methods. The standard 
method (circles), cumulant method of order 2 (squares) and cumulant method of order 3 (triangles) are depicted. The 
12.5\% most central collisions are labeled as central, the centrality 12.5\% -33.5\% as mid-central and 33.5\% -100\% 
as peripheral. For the model calculations the corresponding impact parameters of $b \le 3.4$~fm for central, 
$b=5-9$~fm for mid-central and $b= 9-15$~fm for peripheral collisions have been used
(from Petersen {\it et. al.} \protect{\cite{Petersen}}).} 
\label{Hannah4}
\end{center}
\end{figure}

Contrary, the recent NA49 data at $40 A$~GeV (see
Ref.~\cite{NA49_v2pr40,Petersen} (cf.
Figs. \ref{Hannah3} and \ref{Hannah4}) show a sudden collapse of the proton flow 
for midcentral collisions. At $40 A$~GeV 
this collapse of $v_2$ for protons around midrapidity is
very pronounced while it is not observed at $160 A$~GeV.

Another evidence for the hypothesis of the observation of a first--order phase
transition to QCD is the dramatic collapse of the flow $v_1$ also observed by 
NA49~\cite{NA49_v2pr40}, again around $40 A$~GeV, where the collapse of
$v_2$ has been observed. This is the highest energy at which 
a first-order phase transition can be reached
at central rapidities of relativistic heavy-ion collisions (cf.\
Ref.~\cite{Fodor04,Karsch04} and Fig.~\ref{phasedia}).
Therefore one may conclude that a first-order phase transition at the highest
baryon densities accessible in nature has been seen at these energies
in Pb+Pb collisions. As shown in Ref.~\cite{Paech03}, the
elliptic flow clearly distinguishes between a first-order phase
transition and a cross over.

\section{Summary}

Evidence for a first--order phase transition in baryon--rich dense matter
is recently presented by the collapse of both, $v_1$- and
$v_2$-collective flow of protons from the Pb+Pb collisions at $40 A$~GeV of
the NA49 collaboration. It will soon be possible to study the nature
of this transition and the properties of the QGP at the high-$\mu$/low energy and at the forward
fragmentation region at RHIC and at the future GSI facility FAIR.

This first-order phase transition occurs according to lattice
QCD results~\cite{Fodor04,Karsch04} for chemical potentials above 400 MeV. 
Since the elliptic flow clearly distinguishes between a first-order
phase transition and a cross over \cite{Paech03}, the observed collapse of flow,
as predicted in Ref.~\cite{Hofmann74,Hofmann76}, is a clear signal for
a first-order phase transition at the highest baryon densities.
Calculations from ideal hydrodynamics \cite{Andrade:2006yh} including additional fluctuations
predict an increase of 50\% for fluctuations of the flow; 
however transport models predict an increase by a factor of $2$ 
and $3$ \cite{Vogel:2007yq}. The viscosity coefficient of QGP might
experimentally be determined from these fluctuations.

We predict that the collapse of the proton flow analogous to the
$40 A$~GeV data will be seen in the second--generation experiments at RHIC and FAIR.

\section{Acknowledgements}
We like to thank B. B{\"a}uchle, B. Betz, E. Bratkovskaya, M. Bleicher, A. Dumitru, I. Mishustin, K. Paech, H. Petersen,
D. Rischke, S. Schramm, G. Zeeb, X. Zhu and D. Zschiesche for discussions.


\end{document}